\documentclass[twocolumn,showpacs,preprintnumbers,amsmath,amssymb]{revtex4}

\usepackage{graphicx}
\usepackage{dcolumn}
\usepackage{bm}
\usepackage{color} 

\begin{document}

\title{Thermally activated intersubband scattering and oscillating magnetoresistance in quantum wells}

\author{S. Wiedmann$^{1,2}$, G. M. Gusev$^3$, O. E. Raichev$^4$, A. K.
Bakarov$^5$, and J. C. Portal$^{1,2,6}$}
\affiliation{$^1$LNCMI, UPR 3228, 
CNRS-INSA-UJF-UPS, BP 166, 38042 Grenoble Cedex 9, France} \affiliation{$^2$INSA Toulouse,
31077 Toulouse Cedex 4, France} \affiliation{$^3$Instituto de
F\'{\i}sica da Universidade de S\~ao Paulo, CP 66318 CEP 05315-970,
S\~ao Paulo, SP, Brazil}\affiliation{$^4$Institute of
Semiconductor Physics, NAS of Ukraine, Prospekt Nauki 41, 03028,
Kiev, Ukraine} \affiliation{$^5$Institute of Semiconductor Physics,
Novosibirsk 630090, Russia} \affiliation{$^6$Institut Universitaire
de France, 75005 Paris, France}
\date{\today}

\begin{abstract}
Experimental studies of magnetoresistance in high-mobility wide quantum wells 
reveal oscillations which appear with an increase in temperature to 10~K and 
whose period is close to that of Shubnikov-de Haas oscillations. The observed 
phenomenon is identified as magneto-intersubband oscillations caused by the 
scattering of electrons between two occupied subbands and the third subband 
which becomes occupied as a result of thermal activation. These small-period 
oscillations are less sensitive to thermal suppression than the large-period 
magneto-intersubband oscillations caused by the scattering between the first 
and the second subbands. Theoretical study, based on consideration of electron 
scattering near the edge of the third subband, gives a reasonable explanation 
of our experimental findings.
\end{abstract}


\maketitle

\section{Introduction}

Magnetoresistance oscillations caused by Landau quantization provide important information 
about fundamental properties of electron system in solids. Studies of the Shubnikov-de Haas 
(SdH) oscillations due to sequential passage of Landau levels through the Fermi level \cite{1} 
allow one to investigate the shape of the Fermi surface as well as the scattering processes 
leading to Landau level broadening. Apart from SdH oscillations, there exist oscillating 
phenomena which are not related to the position of Landau levels with respect to the Fermi 
level and, therefore, are less sensitive to temperature. One of the most important examples 
of such phenomena are the magneto-intersubband (MIS) oscillations \cite{2} observed in 
two-dimensional (2D) electron systems with two or more populated dimensional-quantization 
subbands, which are realized in single, double, and triple quantum wells \cite{3,4,5,6,7,8,9,10}.
Recently, magento-oscillations driven by intersubband transitions have
also been reported for 2D electrons on liquid helium \cite{11}. The peculiar magnetoresistance 
properties of 2D electron systems are caused by the possibility of elastic 
(impurity-assisted) scattering of electrons between the subbands. The MIS oscillations 
occur because of a periodic modulation of the probability of intersubband transitions 
by the magnetic field. The maxima of these oscillations correspond to the condition when 
subband splitting energy $\Delta$ is a multiple of the cyclotron energy $\hbar \omega_c$, so 
the Landau levels belonging to different subbands are aligned. Since MIS oscillations 
survive an increase in temperature, they are used to study electron scattering 
mechanisms at elevated temperatures when SdH oscillations completely disappear in the 
region of weak magnetic fields \cite{6,8,9,10}. 

Despite the fact that MIS oscillations are one of the most fundamental manifestations 
of quantum magnetotransport, their properties are not sufficiently studied. In particular, 
the case when one of the subbands is placed close to the Fermi energy and its filling by 
electrons is very small deserves a closer attention. Theoretical studies \cite{12} 
confirmed that MIS oscillations can exist at such small fillings of the upper subband, but 
a detailed experimental investigation of this interesting situation is still missing. From 
the theoretical point of view, it is important to consider scattering mechanisms of 
electrons near the edge of the weakly populated subband and a possibility of probing 
these mechanisms by magnetoresistance measurements.  

We have studied magnetoresistance in symmatric GaAs wide quantum wells (WQWs) with high-mobility 
2D electron gas. Owing to a high electron density and a large well width, these systems form 
a bilayer configuration due to charge redistribution, when two quantum wells near the 
interfaces are separated by an electrostatic potential barrier (see Fig. 1). The presence of 
the two occupied subbands is confirmed by the observation of MIS oscillations in magnetoresistance.
Apart from MIS and SdH oscillations, we observe unusual oscillations which appear when 
temperature is raised to 10~K and persist up to $T=40$~K in the region of magnetic 
fields from 0.35 to 2~T. The period of these oscillations is slightly smaller than the period 
of SdH oscillations. The dependence of resistivity on magnetic field and temperature allows 
us to treat these small-period oscillations as the MIS oscillations caused by electron
scattering between the two lowest subbands and the third subband which is placed slightly 
above the Fermi energy $\varepsilon_F$ (Fig. 1) and becomes populated as a result of thermal 
activation. This conclusion is supported by a theoretical consideration of magnetoresistance, 
which also uncovers the scattering mechanism responsible for Landau level broadening and 
thereby explains the unusual low sensitivity of the small-period MIS oscillations 
to thermal suppression at elevated temperatures. The calculated magnetoresistance is in 
good agreement with the experimental results.

\begin{figure}[ht]
\includegraphics[width=7.cm]{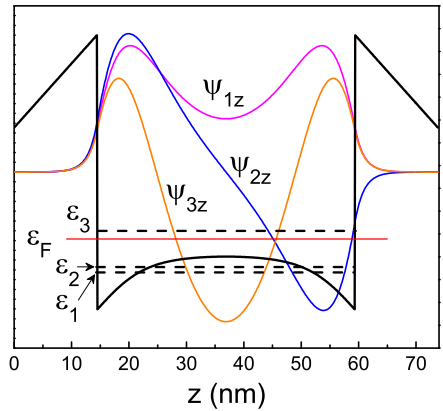}
\caption{\label{fig1}(Color online) Calculated confinement potential profile of our wide 
quantum wells and wave functions of electrons for the first three subbands. Positions 
of the subbands (straight dashed lines) and the Fermi level (straight solid line) are 
schematically shown.}  
\end{figure}

The paper is organized as follows. Section II presents experimental details and 
results. Section III gives theoretical calculation of magnetoresistance and its 
application to analysis and discussion of the experimental data. Concluding remarks 
are given in the last section. The Appendix is devoted to calculation of quantum 
lifetime of electrons in the upper subband.
 
\section{Magnetoresistance measurements}
  
We have studied wide GaAs quantum wells ($w$=45~nm) with an electron density of $n_{s} 
\simeq 9.2 \times 10^{11}$ cm$^{-2}$ and a mobility of $\mu~\simeq 1.9 \times 10^{6}$ cm$^{2}$/V s 
at low temperatures. To achieve both high density and high mobility, the samples have been 
produced according to Ref. \cite{13}, where the barriers surrounding the quantum well are
formed by short-period AlAs/GaAs superlattices. Samples in both Hall bar ($l\times w$=
250~$\mu$m $\times$ 50~$\mu$m) and van der Pauw (size 3~mm $\times$ 3~mm) geometries have 
been studied. The two lowest subbands are separated by the energy $\Delta_{12}=1.40$~meV, 
extracted from MIS oscillation periodicity \cite{6}. This value is in agreement with a 
self-consistent numerical calculation of the electron energy spectrum and wave 
functions (Fig. 1). The small energy separation and the symmetry of the wave functions 
for the two lowest subbands show that the corresponding (symmetric and antisymmetric) 
states are formed as a result of tunnel hybridization of the states in two quantum wells 
near the interfaces. Measurements of the longitudinal resistance $R_{xx}$ have been 
carried out in a perpendicular magnetic field $B$ up to 2.5~T in a cryostat with a
variable temperature insert in the temperature range from 1.4 to 40~K.
As confirmed by theoretical estimates, see Eq. (8) and Fig. \ref{fig6} below, 
the magnetoresistance measurements in the fields below 0.5 T are performed in the regime 
of overlapping Landau levels.

\begin{figure}[ht]
\includegraphics[width=9.4cm]{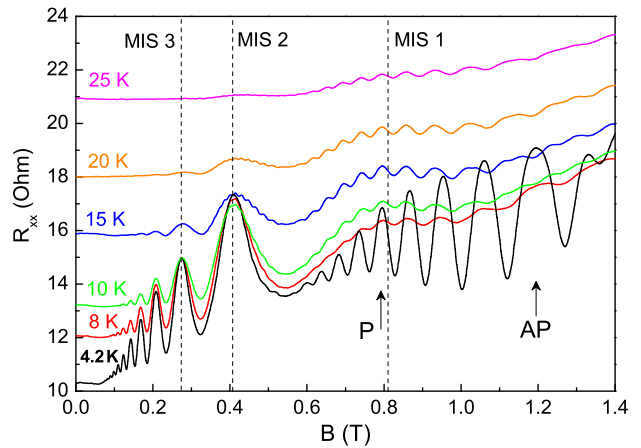}
\caption{\label{fig2}(Color online) (a) Measured longitudinal resistance of a two-subband 
system at different temperatures from 4.2 to 25~K. The vertical dashed lines indicate 
positions of the three main MIS peaks. As temperature increases, the SdH oscillations are 
replaced by a new kind of oscillations which have a smaller period and survive at high 
temperatures. We identify these oscillations as MIS oscillations associated with the 
third subband.}  
\end{figure}

The main results for the magnetoresistance are summarized in Fig. 2. Several 
groups of quantum oscillations, periodic with the inverse magnetic field, are observed. 
At low temperatures we see both SdH (small-period) and MIS (large-period) oscillations,
the latter are caused by electron scattering between the two lowest subbands.
SdH oscillations are visible at 4.2~K in the region of magnetic fields above 0.6~T. 
In this region, SdH oscillations are superimposed on the first MIS peak whose maximum 
is placed at $B \simeq 0.8$~T corresponding to the alignment condition $\hbar \omega_c=
\Delta_{12}$. With increasing temperature, the SdH oscillations are rapidly damped 
and disappear. Disapperance of SdH oscillations for $T> 4.2$~K and in the 
range of magnetic fields studied in the present work can be easily confirmed applying 
the well-known Lifshitz-Kosevich formula containing the specific 
thermal damping factor $(2 \pi^2 T/\hbar \omega_c)/\sinh(2 \pi^2 T/\hbar \omega_c)$.
However, another oscillating pattern is developed at $T \sim 10$~K and 
persists even at $T > 25$~K, when large-period MIS oscillations are strongly damped. The period 
of these high-temperature oscillations is close to the SdH oscillation period but does not 
coincide with it. For example, the SdH peak at $B \simeq 0.8$ T appears to be in phase 
with the high-temperature oscillations (arrow P in Fig. \ref{fig2}), while the SdH peak at 
$B \simeq 1.2$ T stays in antiphase (arrow AP in Fig. \ref{fig2}). At $B \sim 1$ T, the 
high-temperature oscillations replace SdH oscillations in the interval from $T=8$ to 
$T=10$ K. As seen from the plots for 10, 15, and 20~K, the high-temperature oscillations 
also exist in the region of lower magnetic fields corresponding to the second MIS peak. 

\begin{figure}[ht]
\includegraphics[width=9.cm]{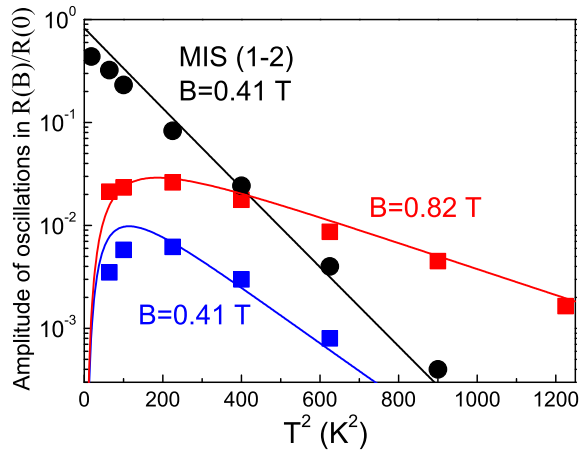}
\caption{\label{fig3}(Color online) Temperature dependence of the amplitudes of 
small-period (squares) and large-period (circles) MIS oscillations. The solid lines 
are the results of theoretical calculation (see Sec. III for details).}  
\end{figure}

Therefore, the nature of the small-period high-temperature oscillations is obviously 
different from the nature of SdH oscillations. The origin of these oscillations can be 
understood if the third subband in our quantum well is included into consideration. 
Indeed, numerical calculations of WQW energy spectrum show that at $T=0$ the third 
subband should be weakly populated at the given electron density. The fact that we do not 
observe the effects associated with the third subband at low temperatures indicates 
a possible error within a few meV in theoretical determination of the position of 
this subband. If we assume that the third subband is placed slightly above the Fermi 
energy ($\varepsilon_3 > \varepsilon_F$) and attribute the small-period oscillations 
to MIS oscillations owing to electron transitions between this subband and the subbands 
1 and 2, the observed properties of these oscillations become clear. First, the 
periodicity in this case is determined by the large splitting energies $\Delta_{13}=
\varepsilon_3-\varepsilon_1$ and $\Delta_{23}=\varepsilon_3-\varepsilon_2$, which 
also explains a slightly smaller period of these oscillations compared to SdH 
oscillations. Next, the thermal activation behavior is explained by the increasing 
number of electrons being able to participate in the intersubband transitions with 
increasing temperature. Finally, the persistence of the oscillations at high 
temperatures follows from their nature: a suppression of MIS oscillations with 
temperature is not related to thermal smearing of the Fermi surface but is 
caused by thermal broadening of Landau levels. 

The problem of temperature behavior deserves a more detailed discussion.
Figure 3 shows temperature dependence of the amplitudes (peak-to-peak values) 
for the small-period MIS oscillations superimposed on the maxima of the first and second 
large-period MIS peaks, $B=0.82$ and $B=0.41$ T. For comparison, a similar dependence 
for the amplitude of the second MIS peak is shown. The plots demonstrate a notable 
difference in behavior for these two kinds of MIS oscillations. While large-period 
MIS oscillations are monotonocally suppressed by temperature, the small-period 
oscillations are characterized by a non-monotonic dependence with a maximum around 
10-15 K and a slower decrease with temperature. In the region of high temperatures, 
all the experimental plots apparently show that the logarithm of the amplitude linearly 
decreases with $T^2$. The slope of this decrease linearly scales with $B^{-1}$, as can 
be seen by comparison of the small-period MIS oscillation amplitudes at 0.41 and 0.82 T.
These data suggest that the suppression of both small-period and large-period MIS 
oscillations is governed by the same mechanism, thermal broadening of Landau levels 
due to enhancement of electron-electron scattering with temperature. Such an effect 
is described in terms of temperature-dependent quantum lifetime of electrons, $\tau(T)$, 
entering the Dingle factor $d$, the latter determines the oscillation amplitude: 
\begin{eqnarray}
d=\exp[-\pi/\omega_c \tau(T)], ~~~~~~ \nonumber \\
\frac{1}{\tau(T)}=\frac{1}{\tau(0)} + \frac{1}{\tau^{ee}(T)},~~\frac{1}{\tau^{ee}(T)}= 
\lambda \frac{T^2}{\hbar \varepsilon_F},  
\end{eqnarray}
where $\tau(0)$ is the quantum lifetime due to elastic scattering. The term $1/\tau^{ee}$, 
where $\lambda$ is a numerical constant on the order of unity, describes the partial 
contribution of electron-electron scattering, which in high-mobility samples dominates 
starting from $T \simeq 10-15$ K. The reliability of Eq. (1) has been proved in numerous 
magnetoresistance experiments; see Refs. \cite{6,8,10,14} and references therein, 
and constant $\lambda$ has been calculated for similar experimental conditions, 
see Refs. \cite{15,16}. It is also worth noting that the Dingle factor in Eq. (1) can be 
different for MIS and SdH oscillation amplitudes. First, owing to the specific energy dependence 
of the electron-electron scattering time, the electron-electron interaction does not suppress 
SdH oscillations \cite{17,18,19}. Next, MIS oscillations are not sensitive to inhomogeneity 
of the electron density in contrast to SdH oscillations \cite{20,21}. In our experiment, 
the quantum lifetime is extracted from MIS oscillations since SdH oscillations are already 
damped in the temperature interval under consideration. 

However, by comparing the slopes of high-temperature suppression of small-period and 
large-period MIS oscillation amplitudes in Fig. 3 at the same magnetic field (0.41 T), it 
is evident that the small-period oscillations are more robust with respect to increasing 
temperature. This interesting and unexpected result may indicate a weaker temperature 
dependence of the Landau level broadening in the third subband compared to subbands 
1 and 2. Such an assumption is confirmed by theoretical calculations. The theoretical 
analysis carried out in the next section demonstrates that the consideration of 
electron scattering near the edge of the third subband explains the whole set of 
the data obtained in our magnetoresistance measurements.
 
\section{Theoretical study}

In the experimentally relevant range of transverse magnetic fields, when the number of Landau 
levels below the Fermi energy is large, the electron transport in 2D systems is conveniently 
described by using either a quantum Boltzmann equation or Kubo formalism based on treatment of 
electron scattering within the self-consistent Born approximation \cite{22,23}. These 
methods are straightforwardly generalized for many-subband systems \cite{24,25,10}. By considering 
the elastic scattering of electrons in the limit of classically strong magnetic fields (when $\omega_c$ is 
much larger than transport scattering rates), one can express the linear dissipative resistivity 
of the electron system with several occupied 2D subbands in the following way \cite{24}: 
\begin{eqnarray}
\rho_d=\frac{m}{e^2 n^2_s} \sum_{jj'} \int d \varepsilon  \left(-\frac{\partial f_{\varepsilon}}{\partial 
\varepsilon} \right) \nonumber \\
\times \frac{k^2_{j \varepsilon}+k^2_{j' \varepsilon}}{4 \pi}  \nu^{tr}_{jj'} (\varepsilon)
D_{j \varepsilon} D_{j' \varepsilon}, 
\end{eqnarray}
where $m$ is the effective electron mass, $e$ is the electron charge, $j$ is the subband index, 
$k_{j \varepsilon}=\sqrt{2 m (\varepsilon-\varepsilon_j)}/\hbar$ is the electron wave number in 
the subband $j$, $\varepsilon_j$ is the subband energy, $f_{\varepsilon}$ is the equilibrium 
(Fermi-Dirac) distribution function of electrons, and $D_{j \varepsilon}$ is the dimensionless 
(normalized to its zero-field value $m/\pi \hbar^2$) density of electron states in the subband $j$. 
The quantity $\nu^{tr}_{jj'} (\varepsilon)$ is defined as
\begin{eqnarray}
\nu^{tr}_{jj'} (\varepsilon)= \frac{m}{\hbar^3} \int_{0}^{2 \pi} \frac{d \theta}{2 \pi}
w_{jj'}[q_{jj'}(\varepsilon)] \frac{q^2_{jj'}(\varepsilon)}{k^2_{j \varepsilon}+k^2_{j' \varepsilon}}, \\
q^2_{jj'}(\varepsilon)=k^2_{j \varepsilon}+k^2_{j' \varepsilon} - 2 k_{j \varepsilon} k_{j' \varepsilon} 
\cos \theta, \nonumber 
\end{eqnarray}
where $w_{jj'}(q)$ is the spatial Fourier transform of the correlators of random scattering potential, $q_{jj'}(\varepsilon)$ is the wave number transferred in elastic collisions and $\theta$ is the scattering 
angle. In many cases, energy dependence of $k_{j}$ and $w_{jj'}(q)$ can be neglected within the 
interval of thermal smearing of the electron distribution. Then $k_{j \varepsilon}$ are taken at 
the Fermi surface, $k_{j \varepsilon}=k_{j \varepsilon_F}=\sqrt{2 \pi n_j}$, where $n_j$ is the sheet 
electron density in the subband $j$, and $\nu^{tr}_{jj'} (\varepsilon)$ are reduced to transport 
scattering rates $\nu^{tr}_{jj'}$ defined, for example, in Ref. \cite{26}.
 
In our samples, where two lowest subbands are closely spaced and almost equally populated while the 
third subband is weakly populated, we have $k^2_1 \simeq k^2_2 \gg k^2_3$. Application of Eq. (2) 
under these conditions gives us the following expression:
\begin{eqnarray}
\rho_d = \frac{m}{2 e^2 n_s} \int d \varepsilon  \left(-\frac{\partial f_{\varepsilon}}{\partial 
\varepsilon} \right) \left[ \nu^{tr}_{11} D^2_{1 \varepsilon} +  \nu^{tr}_{22} 
D^2_{2 \varepsilon} \right. \nonumber \\
\left. + 2 \nu^{tr}_{12} D_{1 \varepsilon}D_{2 \varepsilon} + 
(\nu^{tr}_{13} D_{1 \varepsilon} + \nu^{tr}_{23} D_{2 \varepsilon}) D_{3 \varepsilon} \right].
\end{eqnarray}
The scattering potential in our samples is created mostly by donor impurities localized in 
the side barrier regions, so the correlation between the effective scattering potentials in the 
layers of the double-layer system formed in our WQWs can be neglected. Under this condition, 
see Ref. \cite{24}, the scattering in a symmetric double-layer system is described by equal correlators $w_{11}(q)=w_{22}(q)=w_{12}(q)$. Although this equality is based on a tight-binding description 
of the double-layer system, is holds with good accuracy in our WQWs, as confirmed by calculations 
of the wave functions for the first and second subband. Thus, in the case of closely spaced first 
and second subbands ($k_1 \simeq k_2$), the intrasubband and intersubband rates are almost equal, 
so we use $\nu^{tr}_{11}=\nu^{tr}_{22}=\nu^{tr}_{12} \equiv \nu_{tr}/2$. For the same reasons, 
the scattering rates between the lower subbands and the upper (third) subband are almost independent 
of the lower subband number. Indeed, since $k_{3 \varepsilon} \simeq 0$, one has $\nu^{tr}_{13}=
mw_{13}(k_1)/\hbar^3$ and $\nu^{tr}_{23}= m w_{23}(k_2)/\hbar^3$ with $k_1 \simeq k_2$, while the 
symmetry of the wave functions and the absence of interlayer correlations cause $w_{13}(q) \simeq 
w_{23}(q)$. Therefore, under the approximations valid for our samples Eq. (2) is finally rewritten as
\begin{eqnarray}
\rho_d = \frac{m \nu_{tr}}{e^2 n_s} \int d \varepsilon  \left(-\frac{\partial f_{\varepsilon}}{\partial 
\varepsilon} \right) \left[ D^2_{\varepsilon} +  \eta D_{\varepsilon} D_{3 \varepsilon} \right],
\end{eqnarray}
where $D_{\varepsilon}=(D_{1 \varepsilon}+D_{2\varepsilon})/2$ and $\eta=\nu^{tr}_{13}/\nu_{tr}= \nu^{tr}_{23}/\nu_{tr}$. Notice that the dimensionless parameter $\eta$ is expected to be considerably 
smaller than unity because of a strong suppression of the correlators $w_{jj'}(q)$ at large $q$ in 
the case of scattering by a smooth potential created by remote impurities. 

The first term in Eq. (5), proportional to $D^2_{\varepsilon}$, describes positive magnetoresistance 
with MIS oscillations, typical for double-layer systems \cite{6,27}. The second term is a correction due to 
elastic scattering of electrons between the lower subbands and the third subband. This correction 
is essentially determined by the density of states in the third subband, $D_{3 \varepsilon}$, 
which experience a broadened steplike growth from zero at the edge of this subband, $\varepsilon=\varepsilon_3$. 
Since in our case $\varepsilon_3 > \varepsilon_F$, the third-subband contribution is not essential 
at low temperatures, when $-\partial f_{\varepsilon}/\partial \varepsilon$ is negligible at 
$\varepsilon > \varepsilon_3$. However, when temperature increases and the third subband 
becomes occupied, thermal activation of the elastic scattering between this subband and the two 
lower ones occurs. The second term in Eq. (5) then plays an important role, leading to small-period 
MIS oscillations which we observe in our experiment. As specified above, the magnitude 
of the parameter $\eta$, which determines the difference in the amplitudes between the two kinds of 
MIS oscillations, is affected by the spatial scale of the scattering potential.

To describe the small-period MIS oscillations, it is crucial to consider the density of states 
in the third subband by focusing on the scattering mechanisms which are responsible for its 
broadening and temperature dependence. The density of states $D_{j \varepsilon}$ for 
subband $j$ can be found from the general expression 
\begin{eqnarray}
D_{j \varepsilon} = \frac{\hbar \omega_c}{\pi} \sum_{n=0}^{\infty} {\rm Im} G^{A}_{\varepsilon j n}, \\
G^{A}_{\varepsilon j n}= \frac{1}{\varepsilon -\varepsilon_j-\hbar \omega_c(n+1/2)- 
\Sigma^{A}_{\varepsilon j n}}, \nonumber
\end{eqnarray}
where $G^{A}_{\varepsilon j n}$ is the advanced (index A) Green's function for the electron 
in subband $j$ in the Landau-level representation and $\Sigma^{A}_{\varepsilon j n}$ is the 
corresponding self-energy. The problem of determination of $\Sigma^{A}_{\varepsilon j n}$ in 
the general case is complicated and does not have an exact solution near the subband edge. 
Nevertheless, a physically reasonable result for $j=3$ can be obtained under a simplifying 
approach, when $\Sigma^{A}_{\varepsilon 3 n}$ is replaced by $i \hbar/2 \tau_3$, where 
$\tau_3$ is the quantum lifetime in the subband 3 calculated in the free-electron approximation. 
In clean samples like ours, the main scattering mechanism contributing to the self-energy in the 
important temperature region $T \geq 10$ K is the electron-electron scattering, so we estimate 
$\tau_3$ based on this scattering mechanism, $\tau_3 \simeq \tau^{ee}_3$. The corresponding 
calculation is done in the Appendix and leads to the result
\begin{eqnarray}
\frac{\hbar}{\tau^{ee}_{3}}= \kappa_0(T) \frac{T^{3/2}}{\sqrt{\varepsilon_F}}, 
\end{eqnarray} 
where $\kappa_0(T)$ is a dimensionless function of temperature which depends on the 
form of the wave functions $\psi_{jz}$ shown in Fig. 1. This function is presented 
in Fig. 4, together with the corresponding broadening energy $\hbar/\tau^{ee}_3$ 
according to Eq. (7). For comparison, we also present the quadratic temperature 
dependence of the broadening energy in the lower subbands, $\hbar/\tau^{ee}= 
\lambda T^2/\varepsilon_F$ [see Eq. (1)], caused by electron-electron scattering. 
The constant $\lambda=2.2$ in this expression is determined from experimental data
on thermal suppression of the large-period MIS oscillations in our samples.   

\begin{figure}[ht]
\includegraphics[width=9.cm]{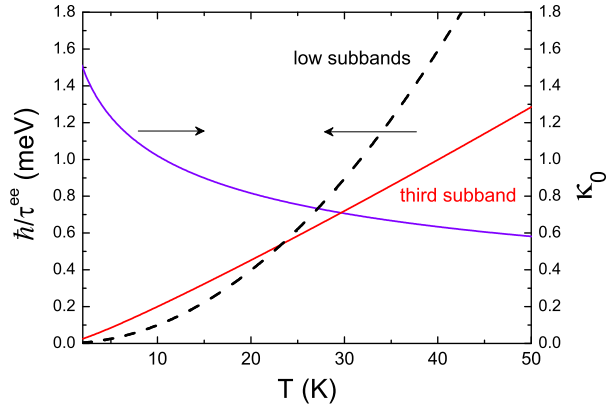}
\caption{\label{fig4}(Color online) Function $\kappa_{0}(T)$ for our WQW system and temperature 
dependence of the inverse quantum lifetime of electrons in the third subband due to electron-electron 
scattering. The dashed line corresponds to the inverse quantum lifetime of electrons in the first 
and second subbands, according to Eq. (1).}  
\end{figure}

Figure 4 demonstrates that temperature dependence of the quantum lifetime of electrons 
in the third subband is weaker than in the first and second subbands. The reason for 
this behavior is rooted in the fact that the third subband is almost empty and contains 
non-degenerate electron gas (see Appendix for details). 

\begin{figure}[ht]
\includegraphics[width=9.cm]{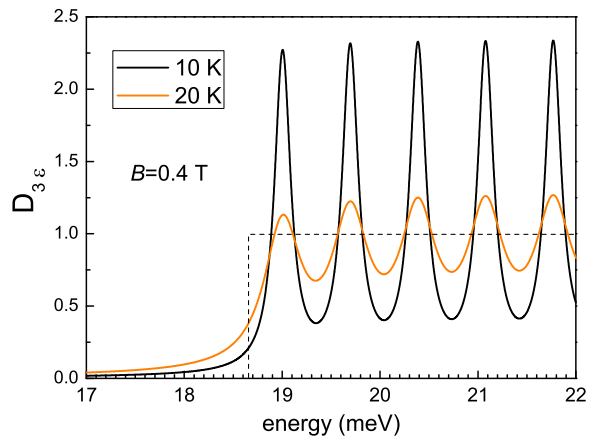}
\caption{\label{fig5}(Color online) Examples of the density of electron states in the third 
subband of our WQW system. The dashed line shows an "ideal" 2D density of states in the 
absence of the magnetic field and collision-induced broadening.}  
\end{figure}

As the self-energy is defined, calculation of the density of states in the subband 3 is 
done according to Eq. (6) by taking the sum over Landau levels $n$ numerically. The result 
of the calculations for $B=0.4$ T and two chosen temperatures is demonstrated in Fig. 5 and 
represents a physically reasonable picture when temperature-dependent energy $\hbar/\tau^{ee}_3$ 
describes broadening of both the Landau levels and subband edge. In spite of the profound 
quantization, the Landau levels are still overlapping, and the oscillating density of 
states at 20 K can be approximated, with a good accuracy, by a harmonic oscillation function.

Next, calculations of the density of states in the two lowest subbands are carried out within 
the self-consistent Born approximation, using the quantum lifetime described by Eq. (1) with 
elastic-scattering contribution $\tau(0)=6.6$ ps determined from low-temperature 
magnetoresistance. Substituting the calculated $D_{\varepsilon}$ and $D_{3 \varepsilon}$ 
into Eq. (5), we finally describe the resistivity. Apart from this numerical calculation, 
it is useful to present an analytical result 
\begin{eqnarray}
\rho_d = \frac{m \nu_{tr}}{e^2 n_s} \left[1+ \eta f_{\varepsilon_3} + d^2 \left(1+ \cos 
\frac{2 \pi \Delta_{12}}{\hbar \omega_c} \right) \right. \nonumber \\
\left. + \eta f_{\varepsilon_3} d d_3 \left( \cos \frac{2 \pi \Delta_{13}}{\hbar \omega_c} +
\cos \frac{2 \pi \Delta_{23}}{\hbar \omega_c} \right) \right],
\end{eqnarray}
based on the approximate (single-harmonic) representation of the densities of states: 
$D_{\varepsilon} \simeq 1-d( \cos[ 2 \pi (\varepsilon-\varepsilon_1)/\hbar \omega] + 
\cos[ 2 \pi (\varepsilon-\varepsilon_2)/\hbar \omega])$ and $D_{3 \varepsilon} 
\simeq 1- 2 d_3 \cos[ 2 \pi (\varepsilon-\varepsilon_3)/\hbar \omega]$ at 
$\varepsilon > \varepsilon_3$. The Dingle factor $d$ is given by Eq. (1), while the 
Dingle factor for the third subband is $d_3=\exp[-\pi/\omega_c \tau^{ee}_3(T)]$.
Equation (8) is valid when $T$ exceeds the broadening energy $\hbar/\tau^{ee}_3$ and 
when $2 \pi^2 T \gg \hbar \omega_c$, so the SdH oscillations are thermally averaged out.
The terms associated with the third subband in Eq. (8) describe small-period 
oscillations due to large subband separation energies $\Delta_{13}$ and $\Delta_{23}$. 
The amplitude of these oscillations is governed, apart from the product of the Dingle 
factors $d d_3$, by a small factor $\eta f_{\varepsilon_3}$. The same factor determines 
a correction to the background (non-oscillating) resistivity. The relative contribution 
of this correction at $T < 30$ K does not exceed 6\% and does not lead to an appreciable 
increase in the resistivity with temperature. The increase in the background resistivity 
observed in experiment, see Fig. \ref{fig2}, is described by thermal enhancement of 
electron scattering by acoustic phonons. 

\begin{figure}[ht]
\includegraphics[width=9.cm]{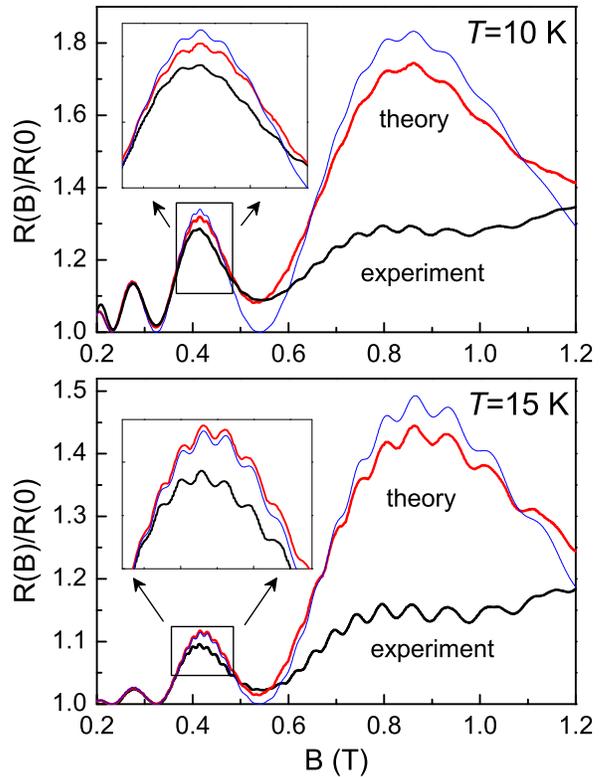}
\caption{\label{fig6}(Color online) Measured and calculated magnetoresistance for our WQW system. 
Two kinds of theoretical plots are shown: based on a numerical calculation (thick line, red) and 
on analytical expression Eq. (8) (narrow line, blue). The regions around 0.4 T are blown up.}  
\end{figure}

The comparison of the measured and calculated resistivity at two chosen temperatures, 10 and 15 K,
is shown in Fig. 6. First of all, by comparing the periodicity of small-period oscillations we 
determine the position of the third subband and find that it is placed at 2.2 meV above the Fermi 
energy (in other words, $\Delta_{13}=19.36$ meV and $\Delta_{23}=17.96$ meV). Below 0.6~T 
the theoretical and experimental plots show a good agreement, while in the region of the first 
MIS peak ($B \sim 0.8$ T) the experiment shows a considerably smaller magnetoresistance than it 
is expected from the theory. This deviation occurs only in the temperature range
we have studied in the present work and is absent at low temperatures ($T < 4.2$~K). We do not 
know exactly the reason for this deviation, probably it is associated with some mechanisms of 
Landau level broadening not taken into account or with the influence of electron-phonon scattering 
on magnetotransport \cite{28}. As concerns the small-period MIS oscillations, the 
best fit of their amplitudes to experimental values is obtained in the whole region of magnetic fields 
by considering $\eta$ as a single adjustable parameter of 
our theory, set at a reasonable value $\eta=0.2$ (a calculation of $\eta$ is possible in 
general, but requires a detailed knowledge of impurity nature and distribution). 
Notice that the results of numerical and analytical methods of calculations are 
in a reasonable accordance, which indicates the reliability of the analytical approach of Eq. (8) 
and validity of the conditions of overlapping Landau levels. Similar 
calculations at the other temperatures also demonstrate a good agreement.  

We also applied the results of our calculations for description of the temperature dependence 
of MIS oscillations of both kinds. As found experimentally, the small-period MIS oscillations 
experience a thermal-activated behavior at low temperatures and a weaker (compared to 
large-period MIS oscillation) thermal suppression at larger temperatures. Although both 
these features are now qualitatively understood, it is instructive to compare directly the 
experimental data on peak-to-peak amplitudes shown in Fig. 3 with corresponding theoretical 
results following from the analytical expression (8). Such theoretical plots are added to 
Fig. 3 and show a good agreement with experiment concerning both non-monotonic temperature 
dependence of the small-period MIS oscillation amplitudes in the region of $T=8-15$ K and 
the decrease in these amplitudes at elevated temperatures. 

\section{Conclusions}

We have designed a WQW structure with high electron density, where two lowest closely spaced 
subbands are occupied by electrons, while the third subband is placed slightly above the 
Fermi energy and, therefore, is not occupied at low temperatures (Fig. 1). By measuring the 
magnetoresistance of this system, we have detected thermally-activated MIS oscillations 
caused by elastic scattering of electrons between the third subband and the two lower subbands. 
These small-period MIS oscillations demonstrate an unusually slow suppression at higher 
temperatures as compared to the well-established temperature dependence of the large-period MIS 
oscillations caused by electron scattering between the lowest (occupied) subbands. Our theoretical 
study has uncovered the reasons for this behavior. The temperature dependence for both kinds 
of MIS oscillations is determined by the influence of electron-electron scattering on the Landau 
level broadening. However, the widely accepted $T^2$ scaling of the broadening energy cannot be 
applied to the case of an almost empty subband, where the carriers have kinetic energies on the 
order of $T$ and form a non-degenerate electron gas near the subband bottom. For such a case, a 
detailed calculation shows that the broadening energy scales with temperature slower than $T^{3/2}$, 
depending on the shape of the wave functions determined by the confinement potential. The theoretical 
dependence of the resistivity on the magnetic field and temperature explains all essential details 
of our experimental results. 

To the best of our knowledge, thermally-activated MIS oscillations have not been reported previously 
in magnetoresistance measurements. In our research, based on a particular structure design, 
we have demonstrated that the existence of such oscillations opens different ways in applications 
of magnetotransport experiments to study electron energy spectrum and scattering mechanisms 
in multi-subband systems. We assume that this research will stimulate further experimental 
and theoretical studies in this direction. 
        
This work was supported by COFECUB-USP (Project number U$_{c}$~109/08), CNPq and FAPESP.

\appendix

\section{Quantum lifetime of electrons in the upper subband}

We start our consideration of electron-electron scattering by neglecting the processes in 
which electrons are transferred between subbands, because such processes require large 
transferred momenta $\hbar {\bf q}$ and, for this reason, are strongly suppressed, 
especially in wide quantum wells. The remaining processes can be viewed as scattering 
of electrons in the subband $j$ by electrons resting either in the same subband or in 
the other subbands. The imaginary part of the self-energy due to electron-electron 
scattering is written, in the momentum representation, as
\begin{eqnarray}
{\rm Im} \Sigma^{(ee) A}_{\varepsilon j {\bf p}} = 2 \pi \sum_{j'} \int \frac{d {\bf p}'}{(2 \pi \hbar)^2}
\int \frac{d {\bf q}}{(2 \pi)^2} U^2_{jj'}(q)  \nonumber \\ 
\times \int d \varepsilon' \int d E  \delta (\varepsilon +E -
\varepsilon_{j {\bf p}+\hbar {\bf q}} ) \delta (\varepsilon' -E -\varepsilon_{j' {\bf p'}-\hbar {\bf q}} ) 
\nonumber \\ 
\times \delta (\varepsilon'-\varepsilon_{j'{\bf p'}} ) \left[f_{\varepsilon'} (1-f_{\varepsilon+E}-f_{\varepsilon'-E} ) + f_{\varepsilon+E} f_{\varepsilon'-E} 
\right],~~
\end{eqnarray}
where $\varepsilon_{j {\bf p}}=\varepsilon_{j}+p^2/2m$, $E$ is the energy transferred in 
collisions, and $U_{jj'}(q)$ is the effective interaction potential:
\begin{eqnarray}
U_{jj'}(q) = \frac{2 \pi e^2}{\epsilon (q+q_0)} I_{jj'}(q),~~~~ \\
I_{jj'}(q)=\int d z \int dz' e^{-q|z-z'|} |\psi_{jz}|^2 |\psi_{j'z'}|^2. \nonumber
\end{eqnarray}
Here $\epsilon$ is the dielectric constant, $q_0=2 e^2 m/\hbar^2 \epsilon$ is the inverse 
screening length, and $\psi_{jz}$ is the envelope wave function for subband $j$. The overlap 
factor $I_{jj'}(q)$ is often set to unity in description of the scattering with small transferred 
momentum which is essential at $T \ll \varepsilon_F$. However, in wide quantum wells (like 
those used in our experiment) this factor becomes important and leads to a significant (2-3 times) 
suppression of the Coulomb interaction, so we keep it in our consideration. To find $I_{jj'}(q)$, 
we use the eigenstates $\psi_{jz}$ obtained in the self-consistent calculation of the energy spectrum 
of our WQW (Fig. 1). Equation (A1) can be derived, for example, from the expressions for 
Green's functions and self-energies presented in Ref. \cite{29}; a generalization of these expressions 
for multi-subband systems is straightforward. Notice that the Green's functions of electrons in 
this derivation are taken in the free-particle approximation, which leads to the appearance of 
the $\delta$-functions of energies in Eq. (A1). 
    
The processes contributing to the self-energy of the third subband, $\Sigma^{(ee)A}_{\varepsilon 3 
{\bf p}}$, include scattering between electrons in the same (third) subband, $j'=j=3$, as well as 
the scattering between electrons in the third and in the lower subbands, $j'=1,2$. Since the 
occupation of the third subband by electrons is low, and $f_{\varepsilon_3} \ll 1$ corresponds to the 
case of non-degenerate electron gas, the latter processes are more significant and are considered 
in the following. For this kind of scattering, the integrals over the variables $\varepsilon'$, ${\bf p}'$, 
and over the angle of the vector ${\bf q}$ in Eq. (A1) can be calculated analytically. Further, expressing 
the quantum lifetime for the third subband according to $1/\tau^{ee}_{3 \varepsilon}= (2/\hbar) {\rm Im} 
\Sigma^{(ee)A}_{\varepsilon 3 {\bf p}} |_{\varepsilon_{3 {\bf p}}=\varepsilon}$, we get
\begin{eqnarray}
\frac{\hbar}{\tau^{ee}_{3 \varepsilon}}= \sum_{j=1,2} \frac{1}{\pi v_{Fj} \sqrt{2m} } 
\int_{-(\varepsilon-\varepsilon_3)}^{\infty} d E \int_{\varepsilon^{(-)}_q}^{\varepsilon^{(+)}_q} d \varepsilon_q
\nonumber \\ 
\times \frac{E}{\sqrt{\varepsilon_q [4\varepsilon_q (\varepsilon-\varepsilon_3)-(E-\varepsilon_q)^2]}} 
\frac{I^2_{3j}(q)}{(1+q/q_0)^2} \nonumber \\ 
\times  \left[f_{\varepsilon+E} + \frac{1}{e^{E/T} -1} \right],  
\end{eqnarray}
where $v_{Fj}$ is the Fermi momentum for one of the lower subbands, $\varepsilon_q=\hbar^2q^2/2m$,
$\varepsilon^{(\pm)}_q=(\sqrt{\varepsilon-\varepsilon_3+E} \pm \sqrt{\varepsilon-\varepsilon_3})^2$ 
Under conditions of our experiment, the difference between $v_{F1}$ and $v_{F2}$ is not essential, so 
$v_{F1}=v_{F2}=\sqrt{2 \varepsilon_F/m}$, where $\varepsilon_F=\hbar^2 \pi n_s/2m$. Next, since 
$f_{\varepsilon_3} \ll 1$, the term proportional to $f_{\varepsilon+E}$ can be neglected 
in Eq. (A3). As a result, the quantum lifetime is represented as   
\begin{eqnarray}
\frac{\hbar}{\tau^{ee}_{3 \varepsilon}}= 
\kappa_{\varepsilon-\varepsilon_3}(T) \frac{T^{3/2}}{\sqrt{\varepsilon_F}}, 
\end{eqnarray}
where $\kappa$ is a dimensionless function of energy and temperature. Numerical 
calculation of $\kappa$ shows that its energy dependence near the edge of the 
third subband ($\varepsilon_3 < \varepsilon < \varepsilon_3+ T$) appears to be weak 
in the important temperature interval from 10 to 30 K. Therefore, below we treat 
the quantum lifetime as energy-independent quantity by taking $\kappa$ at the edge 
of the third subband, $\kappa_{\varepsilon-\varepsilon_3}(T) \simeq \kappa_{0}(T)$. 
Then we obtain the expression
\begin{eqnarray}
\kappa_0(T) = \int_0^{\infty} dx \frac{\sqrt{x}}{e^x-1} F_T(x), \\
F_T(x)=\frac{I^2_{31}(q_x)+I^2_{32}(q_x)}{2 (1+q_x/q_0)^2},~~~q_x=\frac{\sqrt{2m T x}}{\hbar}, \nonumber
\end{eqnarray}
where we introduced a dimensionless variable $x=E/T$. Temperature dependence of $\kappa_0(T)$ 
takes place mostly because of the sensitivity of the squared overlap factors $I^2_{31}$ and 
$I^2_{32}$ to transferred energy $E$ (which, near the edge of the third subband, is directly 
connected to the transferred momentum, $E \simeq \varepsilon_q$). This sensitivity is 
caused by the large width of our quantum well and is essentially determined by the form of 
the wave functions $\psi_{jz}$ shown in Fig. 1. The function $\kappa_0(T)$ calculated according 
to Eq. (A5) is presented in Fig. 4.

In narrow wells, where the characteristic wave number $q$ is small in comparison to the 
inverse well width, the suppression factor $F_T(x)$ in Eq. (A5) can be approximated by 
unity, leading to the universal result $\kappa_0(T) \simeq 2.3$. Then we obtain the 
dependence $1/\tau^{ee}_{3} \propto T^{3/2}$ which is weaker than $1/\tau^{ee} \propto T^{2}$ 
characterizing temperature dependence of quantum lifetime in the lower subbands [Eq. (1)]. 
In wide wells, where the wave number $q$, limited by temperature, may exceed the inverse 
well width, an additional weakening takes place, so the temperature dependence of $1/\tau^{ee}_{3}$ 
calculated for our system and shown in Fig. 4 is different from $T^{3/2}$ and roughly 
resembles a linear function.

\end{document}